\documentclass[12pt,a4paper,english]{article} 

\usepackage{lmodern} 

\usepackage[T1]{fontenc} 
\usepackage[utf8]{inputenc} 
\usepackage{babel} 
\usepackage{geometry} 
\usepackage{setspace} 
\usepackage{pdflscape} 
\usepackage{rotating} 
\usepackage{microtype} 
\usepackage{enumerate} 
\usepackage{natbib} 
\usepackage{floatrow} 
\usepackage{titlesec} 
\usepackage{verbatim} 
\usepackage{caption} 
\usepackage{subcaption} 
\usepackage{endnotes} 
\usepackage{appendix}

\usepackage{amssymb} 
\usepackage{amsmath} 
\usepackage{amsthm} 
\usepackage{bm} 
\usepackage{eucal} 

\usepackage{graphicx} 
\usepackage{epstopdf} 

\usepackage[table,pdftex,dvipsnames]{xcolor} 
\usepackage{array} 
\usepackage{tabularx} 
\usepackage{booktabs} 
\usepackage{longtable} 
\usepackage{threeparttable} 
\usepackage{multirow} 
\usepackage[table-number-alignment=center]{siunitx} 
\usepackage{pdfpages} 
\usepackage[unicode=true,colorlinks=true,citecolor=blue,linkcolor=black,plainpages=false]{hyperref} 

\title{\textbf{Churn Prediction with Sequential Data and Deep Neural Networks}\\ A Comparative Analysis\footnote{All comments, conclusions, and errors are those of the authors only. Please send comments to gary.mena@bdpems.de}}
\author{C. Gary Mena$^1$ \and Arno De Caigny$^{2,3}$ \and Kristof Coussement$^{2,3}$ \and Koen W. De Bock$^4$ \and Stefan Lessmann$^1$}
\date{ $^1$School of Business and Economics, Humboldt University of Berlin \\ $^2$ IÉSEG School of Management \\$^3$LEM - CNRS 9221 \\ $^4$ Audencia Business School  \\[2ex] September 2019}

\begin{document}
\maketitle

\begin{abstract}
Off-the-shelf machine learning algorithms for prediction such as regularized logistic regression cannot exploit the information of time-varying features without previously using an aggregation procedure of such sequential data. However, recurrent neural networks provide an alternative approach by which time-varying features can be readily used for modeling. This paper assesses the performance of neural networks for churn modeling using recency, frequency, and monetary value data from a financial services provider. Results show that RFM variables in combination with LSTM neural networks have larger top-decile lift and expected maximum profit metrics than regularized logistic regression models with commonly-used demographic variables. Moreover, we show that using the fitted probabilities from the LSTM as feature in the logistic regression increases the out-of-sample performance of the latter by 25 percent compared to a model with only static features.

\end{abstract}

\newpage

\section{Introduction}
One of the core tasks in customer relationship management is customer retention. Predicting the probability that a customer will churn has an important role in the design and implementation of customer retention strategies, especially in saturated industries like the financial or telecommunications sectors. One of the reasons is that in such industries the potential customer base of the (relevant) market is close to be fully allocated between the different competitors. Therefore, the value associated with customer retention tends to be larger than the value obtained from acquiring new customers, which in turn fosters the development of churn management strategies~\cite{HADDEN20072902}. 


The improvement of the predictive performance of churn models is important for targeting and the design of marketing strategies that aim to reduce churn. With the increase in computational power and new data sources, deep learning methods have exploited previously unused features such as social network (\citet{OSKARSDOTTIR2017204}) or textual information (\citet{DECAIGNY2019}) to enhance the predictive performance of customer churn models. However, sequential features are an important group of features that cannot be readily incorporated in off-the-shelf machine learning algorithms without suitable transformations. For example,~\cite{dirmarkt2019} use moving averages as well as recent lagged values of sequential data as features for random forest and neural networks. Although there are attempts to incorporate sequential information directly in churn prediction models as in~\cite{CHEN2012461}, the documented performance of recurrent neural networks provides an alternative approach that can enhance the predictive power of churn models by directly exploiting the sequential nature of behavioral features.

In direct marketing, for example, three sequential variables that can impact the predictive power of churn models are recency, frequency, and monetary  variables (henceforth RFM variables). As defined in general terms in~\cite{zhang2014predicting}, recency is the time period since the customer’s last purchase to the time of data collection, frequency is the number of purchases made by individual customers within a specified time period, and the monetary value variable represents the amount of money a customer spent during a specified time period. Intuitively, in contractual markets RFM variables help to characterize the relative behavior of customers with the firm over time. Thus, these variables help to determine which customer are more prone to churn since the customer might alter her behavior during the period when she is about to churn in contrast to non-churning customers.

The main questions that we aim to answer in this paper are i) what is the relative performance of LSTM models with RFM variables compared to off-the-shelf models with static features, and ii) what is the best way to incorporate sequential information into static models. Accordingly, we contribute to the empirical churn modeling literature by assessing the predictive performance of Long-Short Term Memory architectures to model RFM variables and by providing an empirical application using data from an European provider of financial services. Concretely, we show that i) the predictive performance of RNNs and RFM data is higher than the performance of regularized logistic regression without RFM data, and ii) using RNNs to summarize the information contained in RFM variables is an effective alternative to incorporate the latter in off-the-shelf machine learning methods. 

The document is organized as follows. In section~\ref{sec:litrev} we provide a succinct literature review. In section~\ref{sec:dataexp} we describe the data and our experimental approach. Section~\ref{sec:results} reports the main results of the document, and the final section concludes.

\section{Literature Review}
\label{sec:litrev}
The literature on empirical churn modeling in static settings using cross-sections of data is well studied and~\cite{VERBEKE2012211} provide an extensive benchmark of classification techniques using several real-life data sets. Alternative techniques to analyze churn modeling includes survival analysis (\cite{VANDENPOEL2004196}) however we focus on classification analysis given the data that we have available. 

When there are time-varying features then different aggregation procedure are available so that their information can be used with machine learning classification methods (see~\citet{WEI2002103}, and~\citet{Songtimevar}). The reason why they cannot be directly used is that a majority of the classification methods requires one observation per customer but when there are time-varying features one can usually follow the behavior of the same customer over time and the estimation classification methods cannot directly exploit this type of information. One important exception is the work of~\cite{CHEN2012461} who explicitly consider and exploit the information from longitudinal data for customer churn prediction. Specifically, they propose a novel framework called hierarchical multiple kernel support vector machine that without transformation of time-varying features improves the performance of customer churn prediction compared to SVM and other classification algorithms in terms of AUC and Lift using data sets from the Telecom, and other non-financial industries.

The rising popularity of deep neural networks methods for sequential data has fostered an increase in their applications to churn modeling as the overview in Table~\ref{tab:refsann} shows. To synthesize the results from this literature,~\citet{Martins2017} shows that the performance of LSTM taking into account the time-varying features performs as well as aggregating this information using their average and a random forest algorithm.~\citet{Tan2018} propose a network architecture that combines CNN and LSTM networks that outperforms them individually as well as other algorithms that do not use sequential data in terms of AUC, precision-recall, F1-score, and Mathews correlation coefficient.~\cite{Wangperawong2016} process the time-varying features such that they can be used as images and then apply a CNN architecture, but offer no comparative performance of their approach. In a similar vein,~\cite{Zaratiegui2015} encodes the sequential information as images and then applies a CNN and shows that after encoding the CNN performs better than random forests and extreme gradient boost classifiers. Finally, the study of~\cite{Zhou2019} combines different network architectures to leverage the sequential data and show that this combination outperforms CNN, LSTM and classifiers that do not use the time-varying information like random forest and extreme gradient boosting. 

\begin{table}[h]    \centering \small
    \caption{Previous Studies on Customer Churn Prediction Using Longitudinal Data and Neural Networks}
    \label{tab:refsann}
\resizebox{\textwidth}{!}{
    \begin{tabular}{llllll}
\toprule    
    Study     & Neural Network Architecture & Sample Size (customers) & Industry & Type  &  Sequential Features \\
\midrule    
Tan et al. (2018) &  BLA ( = LSTM + CNN) & $120x10^3$ and $156x10^3$  & MOOC, Online Services & IECD proceeding & Subscription characteristics \\
Wangperawong et al. (2016) & CNN & $1x10^6$ & Telecom & Working Paper & Data, voice, sms usage \\
Zaratiegui et al. (2015) & CNN & $132x10^3$ & Telecom & Working Paper  & Call record, Topup \\
Zhou et al. (2019) & DLCNN (= LSTM + CNN) &  $1x10^6$ & Music Streaming & Conf. Proceedings & Transaction and log activity \\
\bottomrule
    \end{tabular}
}
\end{table}

To conclude this section, note that the last three columns show that the previous works that look to improve the performance of churn models using time-varying information is still in an early stage. Moreover, different industries offer different types of time-varying features which a priori does not tell us whether incorporating such type of information leads to a higher performance of the models.  Thus, this paper makes a contribution by assessing the performance of different classification algorithms and alternative  ways of incorporating the time-varying features in the financial services industry. Specifically, we use recency, frequency, and monetary value longitudinal information, in addition to demographic data, to evaluate the prediction of recurrent neural networks as well as the well-established logistic regression that require aggregating the information for training the model.

\section{Data and Experimental Setup}
\label{sec:dataexp}

\subsection{Data: Definitions and Processing}
An important step in our analysis is to choose a suitable definition of the target variable. The information available in the data set makes it possible to construct churn measures using data related to contract closure of all products.\footnote{Customers can have access to four different type of financial products. To preserve the anonymity of the financial institution we refrain from mentioning the names of the products.} Concretely, the observation period that we use to determine whether a customer churns starts in 01/04/2018 and we use a time window of 12 months.\footnote{A 12 month time window is also used in~\cite{BUREZ20094626} for one of their datasets.} Thus, a customer is labeled as having churned if she closes all contracts during 01/04/2018 and 01/04/2019. Although we could define churn using a six month period, the number of people who actually churned would be greatly decreased which could potentially impact the predictive performance of the estimators. Moreover, a few customers are characterized as churners using the six month definition but are clients again under the 12 month definition, which implies that by using a 12 month the observed churn behavior is not a temporary phenomenon but rather permanent.

Regarding the features, the dataset contains i) demographic variables such as age, gender, social and marital status, ii) customer-company characteristics such as length of relationship, and iii) variables related to customer behavior other than RFM variables like customer-company communication information. In the raw data, features other than RFM data were already normalized in the 0-1 interval and we refer the interested reader to~\cite{DECAIGNY2019} for details about processing of missing values and outliers. Note that we constrain the observations to customers that have available RFM and target information for each of the 36 months before the 12 month observation period, and focus on the treatment of time-varying features.

While recurrent and convolutional neural networks can make use of time-varying features, off-the-shelf methods like regularized logistic regression or random forests require aggregating the time-varying information into a single observation. Hence, the treatment of such variables plays an important role behind the results of the experiments. One option to aggregate the time-varying variables is taking their average value per customer, expecting that differences in the levels between churners and non-churners is salient after the transform. A second transformation is by taking the average of first differences of the sequential data. Assuming the data is sorted by customer and date, this effectively reduces to taking the difference between the last and first observation for each customer and dividing by the number of available periods\footnote{For example, focusing on the numerator the following holds: $(x_{t} - x_{t-1}) + (x_{t-1} - x_{t-2}) = x_{t} - x_{t-2} $}. For our final aggregation procedure we use the last six values of each RFM feature, normalized by its average value of the last quarter.  The motivation is that departures from the average behavior are more important than the levels of the observed RFM data. Notice that this average value is not estimated as a moving average. 

\begin{figure}[h!]\centering
	\caption{Mean RFM data by churn status over time} \label{fig:rfmstat}
	\begin{subfigure}{0.32\textwidth}
		\subcaption{Frequency}
		\includegraphics[width=\linewidth, height=5cm]{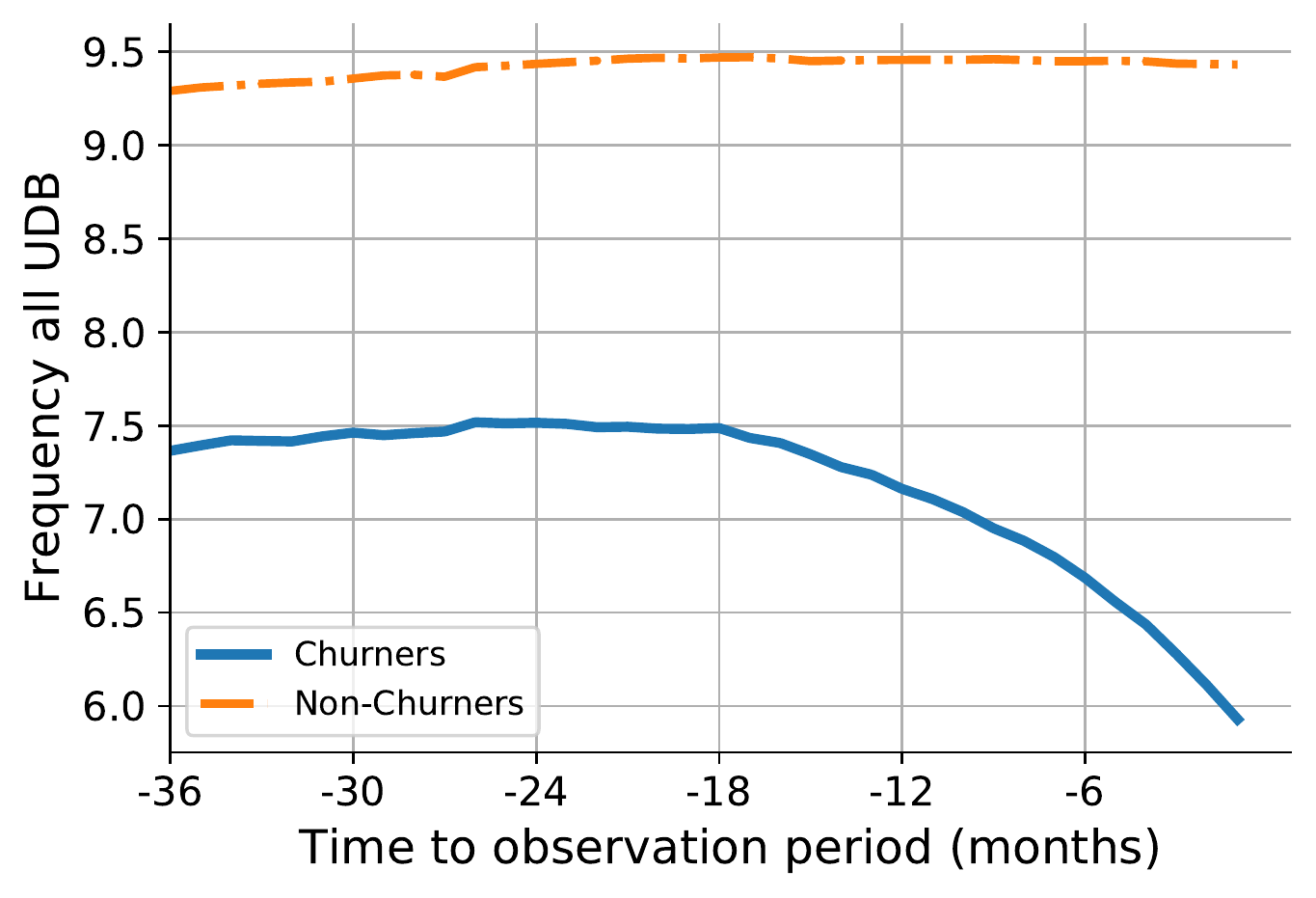}
	\end{subfigure} %
	\begin{subfigure}{0.32\textwidth}
		\subcaption{Recency}
		\includegraphics[width=\linewidth, height=5cm]{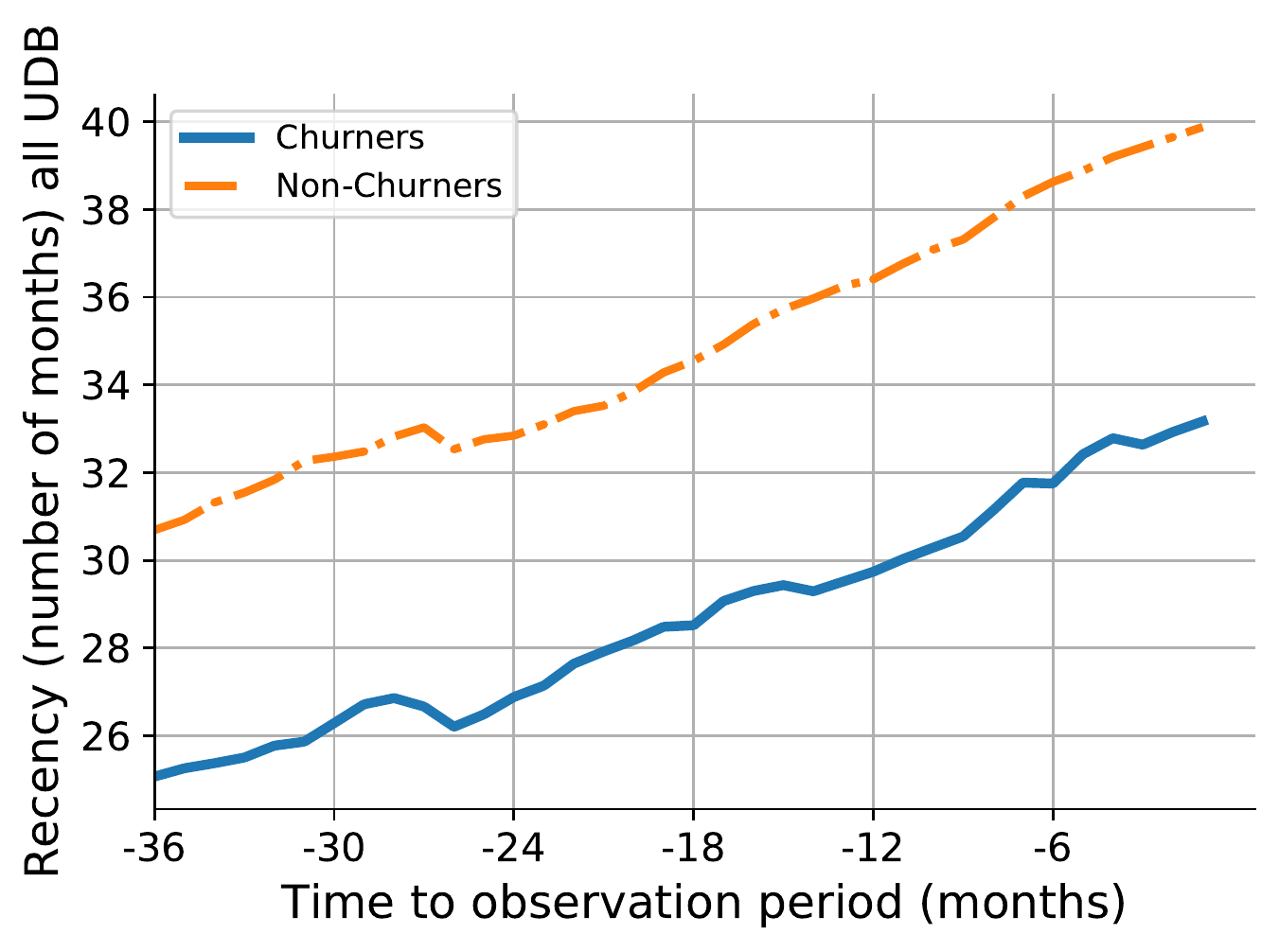}
	\end{subfigure}%
	\begin{subfigure}{0.32\textwidth}
		\subcaption{Monetary value}
		\includegraphics[width=\linewidth, height=5cm]{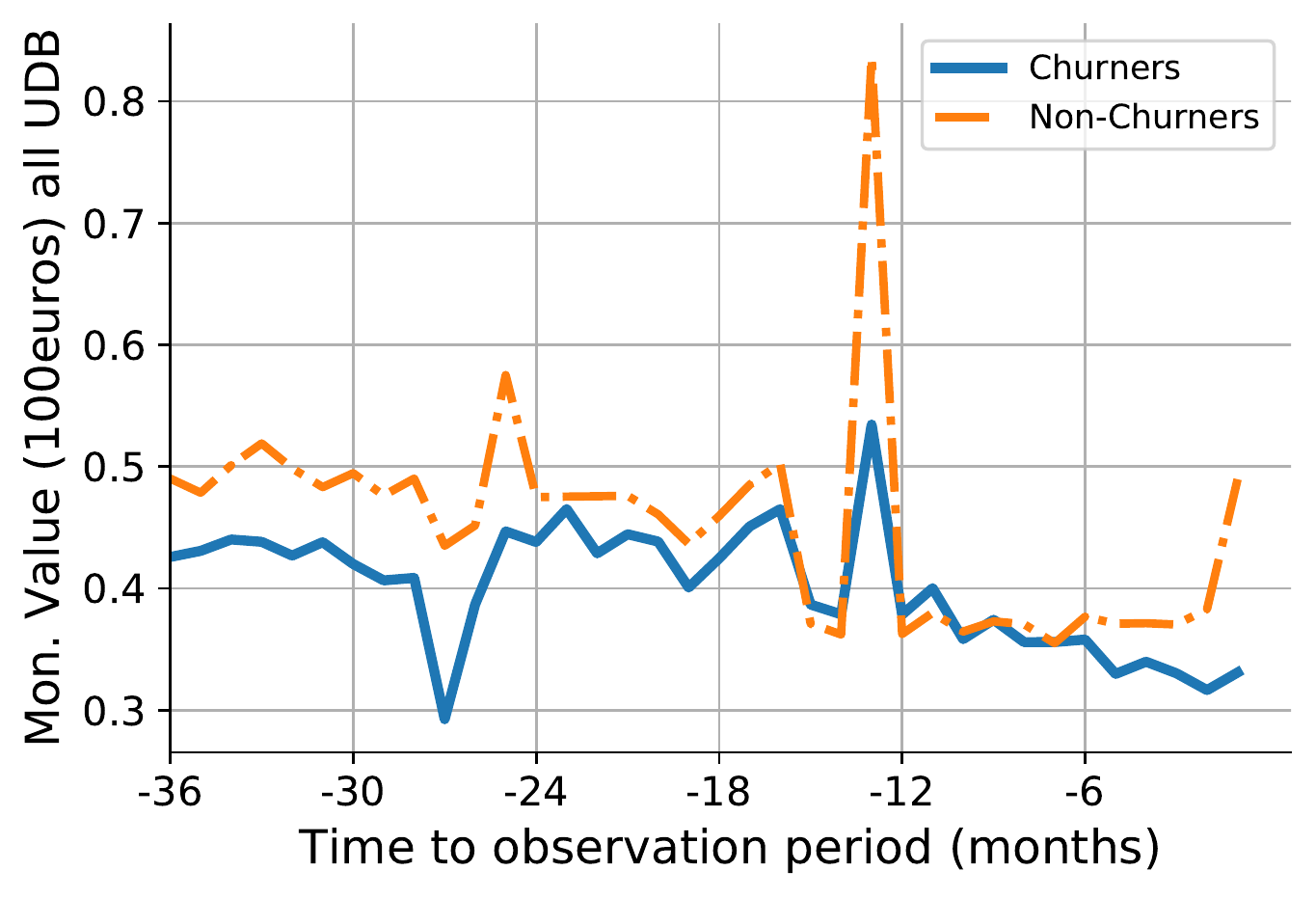}
	\end{subfigure}
	\vspace{-10pt}
	\floatfoot{\footnotesize{\textit{Source:} Own calculations. \\ \textit{Notes:} Sample includes only customers who where observed during all the 36 months prior to the target observation window. }}
\end{figure}

Figure~\ref{fig:rfmstat} presents descriptive statistics of the mean RFM data evolution over time by type of customer. Panel (a) clearly shows that the average value of the frequency feature decreases for churners as we approach the observation window. Such behavior of the feature should be well captured by our aggregation procedures and thus provide a good comparison when compared to the results without aggregation from the neural networks. Panel (b) reveals that churners have on average a lower recency value compared to non-churners. Finally, panel (c) shows no substantial differences of the average monetary value for churners and non-churners beyond a decreasing trend  for churners during the months close to the observation period.

\begin{figure}[h!]\centering
	\caption{Mean of RFM data relative to the mean value of the previous three months  by churn status over time} \label{fig:rfmstat2}
	\begin{subfigure}{0.32\textwidth}
		\subcaption{Frequency}
		\includegraphics[width=\linewidth, height=5cm]{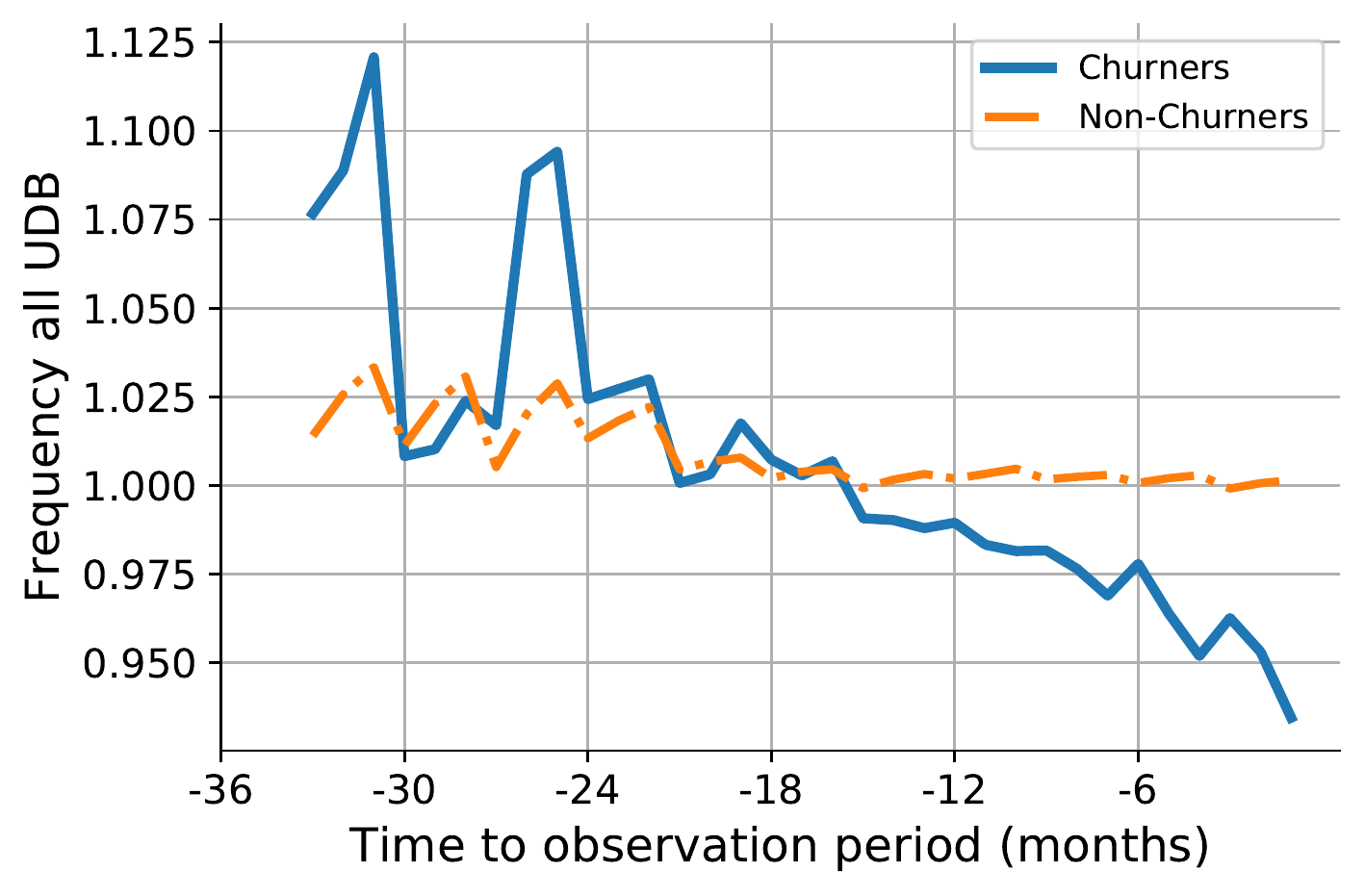}
	\end{subfigure} %
	\begin{subfigure}{0.32\textwidth}
		\subcaption{Recency}
		\includegraphics[width=\linewidth, height=5cm]{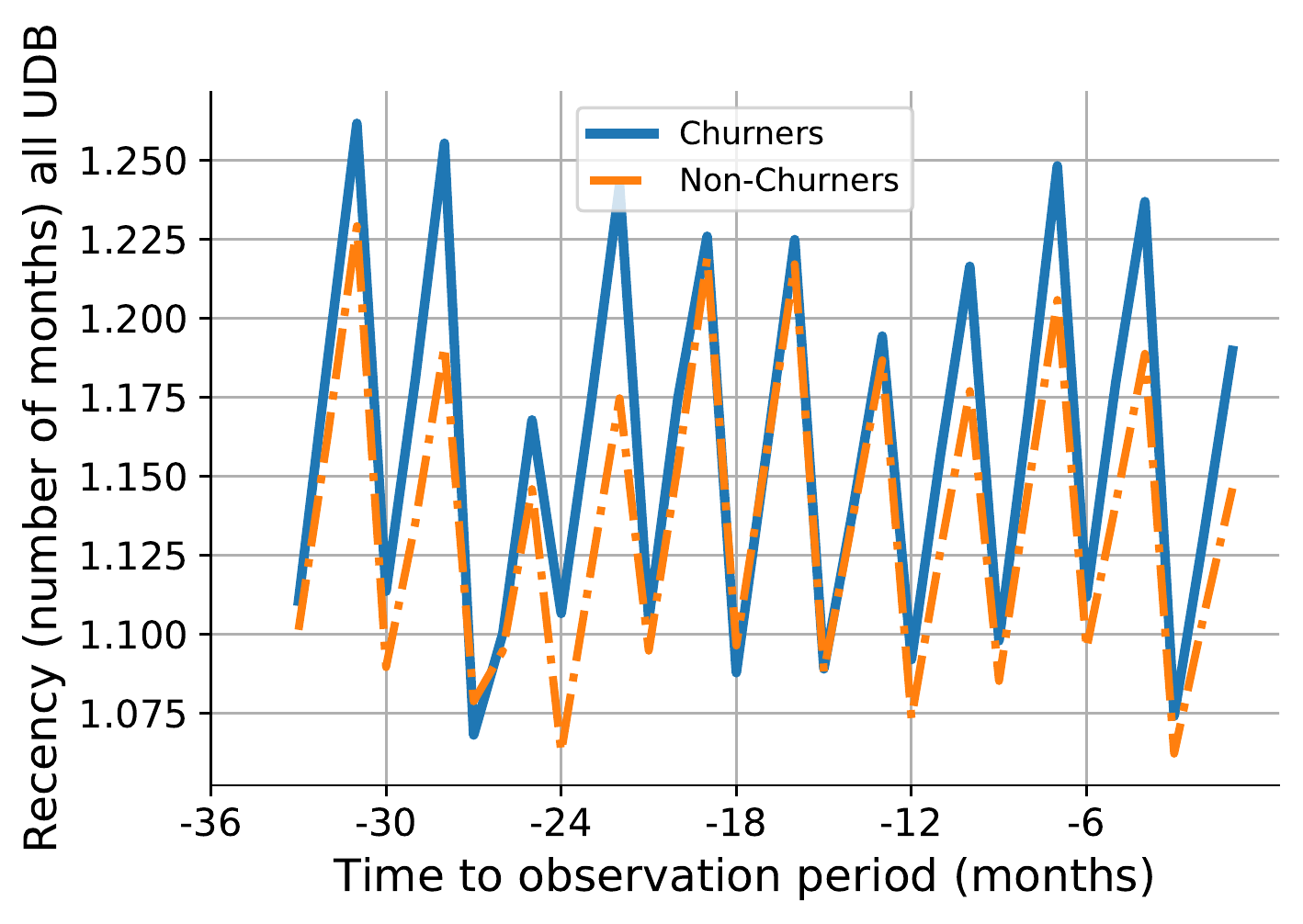}
	\end{subfigure}%
	\begin{subfigure}{0.32\textwidth}
		\subcaption{Monetary value}
		\includegraphics[width=\linewidth, height=5cm]{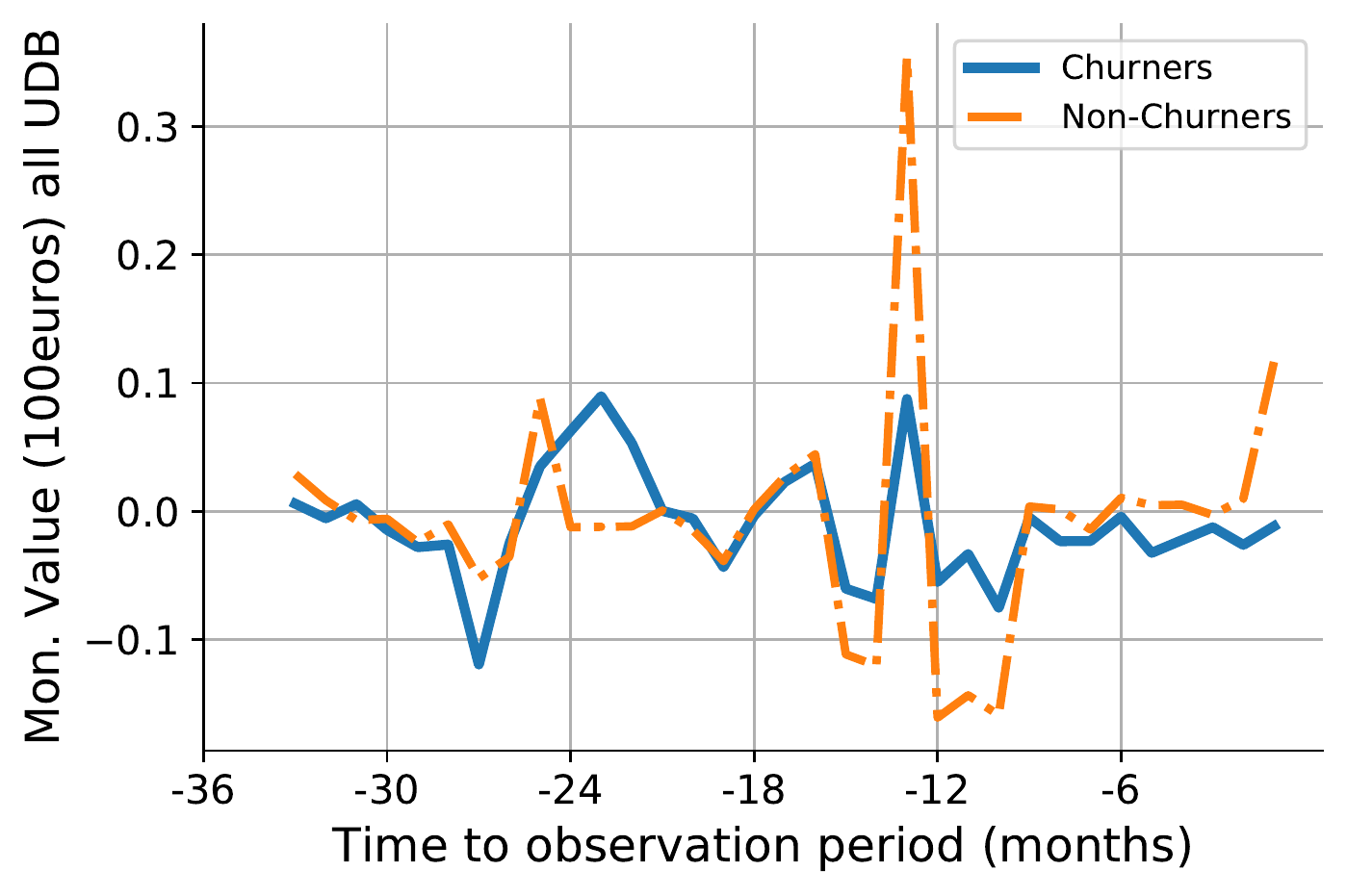}
	\end{subfigure}
	\vspace{-10pt}
	\floatfoot{\footnotesize{\textit{Source:} Own calculations. \\ \textit{Notes:} Sample includes only customers who where observed during all the 36 months prior to the target observation window. To obtain the mean for each customer the RFM is divided by the average of the RFM indicator of the previous quarter. For example, for months 10, 11, and 12 the, say, recency indicator is divided by the average of recency during months 7, 8, and 9, and then the mean is estimated over all customers for a given date.}}
\end{figure}

Figure~\ref{fig:rfmstat2} presents the average values of RFM variables when they are normalized by its average value of the last quarter of the year. In this case, panel (a) is the only one where there are clear differences between the descriptive statistics of churners and non-churners. The figures also highlight how due to the normalization the average difference in levels between groups disappears and only differences in trends remain.

\subsection{Experimental Design}
The main question that we aim to answer with the experimental design is i) what is the relative performance of LSTM models with RFM variables compared to standard models with static features, and ii) what is the best way to incorporate sequential information into static models. In this subsection we document our experimental design in terms of the chosen algorithms, hyper-parameter tuning strategy, handling of class imbalance, and evaluation metrics.

We choose to use the regularized logistic regression as a baseline model. This model is regularly used in industry because it can offer an interpretable measure of the effect of the features on the predicted probabilities of churn. Furthermore, one needs to tune only one hyper-parameter which is the regularization term. We use a Lasso-type of regularization and the set of considered hyper-parameters are given in table~\ref{tab:metapar1}.

\begin{table}[!htbp]\centering
	\caption{Hyper Parameters Considered for Non-sequential Models} \label{tab:metapar1}
	\resizebox{0.95\textwidth}{!}{	
\begin{tabular}{l l l l }
\toprule
Learner& Meta parameter & Broad Tune \\
\midrule
Regularized Logistic Reg. & l-measure & l1 (lasso)    \\ 
& regularization C &  [0.001, 0.01, 0.02, 0.03, 0.04,  0.05, 0.1, 1] \\ 
\bottomrule
\end{tabular} 
} 
	\floatfoot{\footnotesize{\textit{Notes:} All algorithms implemented in Python. Intercept is not penalized.}}
	\vspace{-10pt}
\end{table}

Regarding the neural networks, we choose a LSTM architecture to work with the RFM variables. Table~\ref{tab:metaparDNN1} presents the set of hyper-parameters to tune and the considered values for the search.

\begin{table}[!htbp]\centering
	\caption{Parameter Setting for Experiments with Deep Neural Networks} \label{tab:metaparDNN1}
	\resizebox{0.95\textwidth}{!}{	
\begin{tabular}{l l l l l l l l}
\toprule
Architecture & Layers & Hidden Units & Filter Size & Optimizer & Learning rate & Epochs & Batch size \\
\midrule
LSTM & 1 & [5, 10, 25, 30]  & da & adam  & [0.001] & [10, 25, 50, 75] & [10, 25, 50, 100, 250] \\
\bottomrule
\end{tabular} 
} 
	\floatfoot{\footnotesize{\textit{Notes:} Algorithms are implemented with Keras using Tensorflow as backend. da = does not apply}}
	\vspace{-10pt}
\end{table}

To tune the hyper-parameters of the algorithms we use a nested cross-validation procedure. We choose to work with three outer folds, and four inner folds. We use the inner folds to tune the hyper-parameters (based on the AUC metric), and use the outer folds to compute the full set of evaluation metrics. Notice that after running the inner cross-validation there is the chance of obtaining different sets of tuning parameters for each model, where a model is defined based on the algorithm and the features used. For instance, since there are three outer folds, there can be up to 3 set of parameters for any model after running the four-fold inner cross-validations. In the case that these three sets of hyper-parameters are non-repeated we report the results of the model with the highest AUC metric. When the set of hyper-parameters is repeated we report the evaluation metrics for the model that appeared more times. In the best case scenario the three set of tuning parameters for any given model coincide.

Since we also consider using the LSTM fitted probabilities as feature in the logistic model, we need to take measures to reduce over-fitting. To do so, we divide the training sample of each inner and outer fold into k-folds. Then we train the LSTM with the optimal set of hyper-parameters on k-1 folds and estimate the probability in the fold that was left out. This way, the fitted probability from the LSTM is free from leaking information. For the test folds we use the corresponding training fold to train the LSTM model and then obtain the fitted probabilities.

To evaluate the performance of the models we report the area under the receiver operating curve (AUC), top-decile lift, and expected maximum profit.\footnote{The code for the expected maximum profit is taken from \url{https://github.com/estripling/proflogit-python/blob/master/proflogit/empc.py}. We take the customer lifetime value to estimate the expected maximum profit based on the values in~\cite{DECAIGNY2019}, who use this information to compute the profit of a customer retention campaign.} Each metric is based on a different rationale. The AUC is a common metric applied in the literature that offers a performance measure of a classifier but it does not take into account misclassification costs, which are important from a marketing perspective. Top-decile lift, on the other hand, allows to evaluate the performance of a model within the top ten percent of customers with the highest probability of attrition which is useful when we consider that marketing campaigns are meant to be targeted, but is still not a profit-based metric for evaluation. Thus, by measuring the expected maximum profit we can evaluate the performance of the models from a marketing-relevant perspective since this metric is defined as the expected maximum profit of a classifier with respect to the distribution of classification costs (\citet{VERBEKE2013211}).

One of the most salient stylized facts when modeling churn is that churners tend to represent a small proportion of the observed sample. To illustrate this, in the full training data set the ratio of churners to non-churners is 938 / 384733, which translates to a 0.243 percent of churners. This imbalance in the target variable can have a detrimental effect in the performance of the algorithms if not handled appropriately.~\cite{BUREZ20094626} analyze the problem of class imbalance and show that under-sampling leads to better accuracy compared to more complex sampling techniques. Based on this, we choose an under-sampling approach. Given the high number of observations in the data we do not expect under-sampling to introduce a problem in the estimation due to a decreased sample size. This will also help us to reduce the training time of the algorithms. The under-sampling strategy consists of obtaining two observations from the non-churners sample at random for each customer that churned. Note that we apply this strategy only after the data is split for the nested cross-validation. Finally, features in the training data set are always standardized. The parameters for the standardization are obtained from the training data before under-sampling and then applied to the corresponding test set. 




\section{Results}
\label{sec:results}
Table~\ref{tab:results} reports our main empirical results. The columns present the average value of the respective evaluation metric over the outer folds, whereas the rows show each of the models that we consider. The first row shows that a regularized logistic regression without RFM information and only static features performs the worst compared to the other models according to top-decile lift and EMPC measures. A salient feature is that the predictive performance of the LSTM, which only has RFM features, is higher than that of the logistic model when we use top-decile and EMPC as evaluation metrics. 

When we use the LSTM to extract the probability of churning and use this probability as feature in the logistic model with static features (row 5) we obtain better results for each evaluation metric as compared to the logistic model with aggregated RFM features (rows 2, 3, and 4). This is the main result of the experiment since it provides evidence that using raw RFM data that is commonly available in the financial sector helps to improve the performance of the model without relying on further aggregation procedures. This result also highlights the importance of including RFM information in this study case where the firm operates in the sector of financial services because the magnitude of the improvement in some metrics is not marginal. For example, note that the logistic model with LSTM probabilities has a top-decile lift metric of 4.211 which means that we improve by this amount the number of identified churners as compared with randomly selecting ten percent of the sample. This represents a 25.7 percent\footnote{=(4.211 - 3.350) / 3.350 * 100} improvement  for the lift metric compared to the logistic model without LSTM probabilities. Furthermore,  the logistic model with LSTM probabilities has an EMPC measure that is three times larger than a model with only static features.

The next group of rows (6, 7, 8) uses the LSTM fitted probabilities and the aggregated RFM features in the logistic model. This allows us to assess whether the aggregated time-varying features still help to enhance the performance of the logistic model after we summarized the RFM information through the probability from the LSTM. Results show that the lift metric in these rows is not higher than that of row (5). Similarly, the EMPC metric for these models is at most as large as that of row (5). Thus, this indicates that the fitted probabilities from the LSTM summarizes well the information of the time-varying features. Since we use many aggregations of the RFM variables, this is an important result because it tells us that instead of having to decide which of the aggregations to use, one could just rely on a single summary measure such as the fitted probability from the LSTM. 

\begin{table}[!htbp]\centering
\caption{Mean evaluation metrics} \label{tab:results}
\begin{tabular}{lrrr}
\toprule
Model &    AUC &   Lift &   EMPC \\
\midrule
Only static                                       &  0.746 &  3.350 &  0.004 \\
Static + agg. RFM                                 &  0.768 &  3.964 &  0.006 \\
Static + norm. lagged RFM                         &  0.763 &  3.913 &  0.007 \\
Static + agg. RFM + norm. lagged RFM              &  0.774 &  4.083 &  0.009 \\
Static + LSTM prob.                               &  0.775 &  \textbf{4.211} &  \textbf{0.012} \\
Static + LSTM prob. + norm. lagged RFM            &  0.770 &  4.160 &  0.012 \\
Static + LSTM prob. + agg. RFM                    &  \textbf{0.779} &  4.186 &  0.009 \\
Static + LSTM prob. + agg. RFM + norm. lagged RFM &  0.774 &  4.177 &  0.009 \\
LSTM Neural network                               &  0.741 &  4.101 &  0.006 \\
\bottomrule
\end{tabular}

	\floatfoot{\footnotesize{\textit{Notes:} Evaluation metrics estimated as the mean over the 3-outer folds based on cross-validation. Lift refers to top decile lift, while EMPC refers to the Expected Maximum Profit Measure for Customer Churn (EMPC). All model results are based on l1-regularized logistic regression, except the LSTM neural network which uses the RFM sequential data. Static features refers to customer demographics, customer behavior, and customer contact variables}}
	\vspace{-10pt}
\end{table}

\section{Final Comments}
In this document we assess the predictive performance of neural network architectures for sequential data using RFM variables from a provider of financial services. We use the LSTM model to predict the probability of churning and evaluate its performance against the results of regularized logistic model. Results show that top-decile lift and EMPC measures of the LSTM model with RFM data are higher than the values of the logistic model with only static features. Moreover, when we use the LSTM fitted probabilities and standard static features with the logistic model we obtain the best results as measured by lift and EMPC.

Our results have important implications for churn modeling in the specific case of the financial industry because RFM data is likely to be readily available in this sector and its inclusion in predictive churn models is facilitated through deep learning model such as the LSTM as we show in this paper. This also highlights the importance of incorporating different types of dynamic behavioral data in churn modeling in combination with deep learning methods, which is an open area for future research. 

While the use of deep learning models comes at the cost of increasing the number of tuning parameters and in some cases of training time, we argue that this limitations will be less constraining with the continue increase of computational power.

\newpage
\appendix
\appendixpage

\begin{figure}[h!]\centering
	\caption{Lift curves for LSTM and Logistic models} \label{fig:lift}
	\begin{subfigure}{0.32\textwidth}
		\subcaption{Outer Fold 1}
		\includegraphics[width=\linewidth]{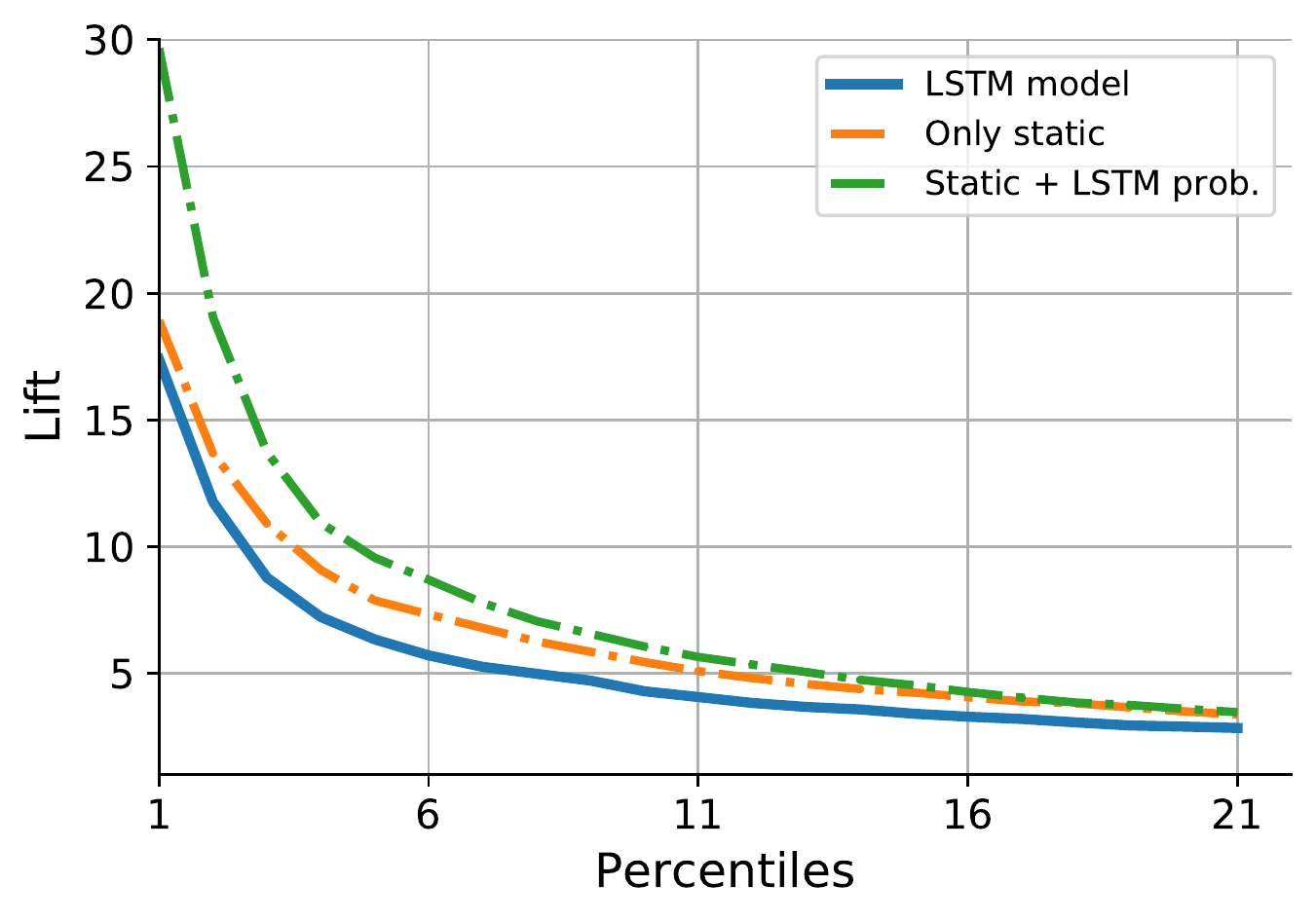}
	\end{subfigure} %
	\begin{subfigure}{0.32\textwidth}
		\subcaption{Outer Fold 2}
		\includegraphics[width=\linewidth]{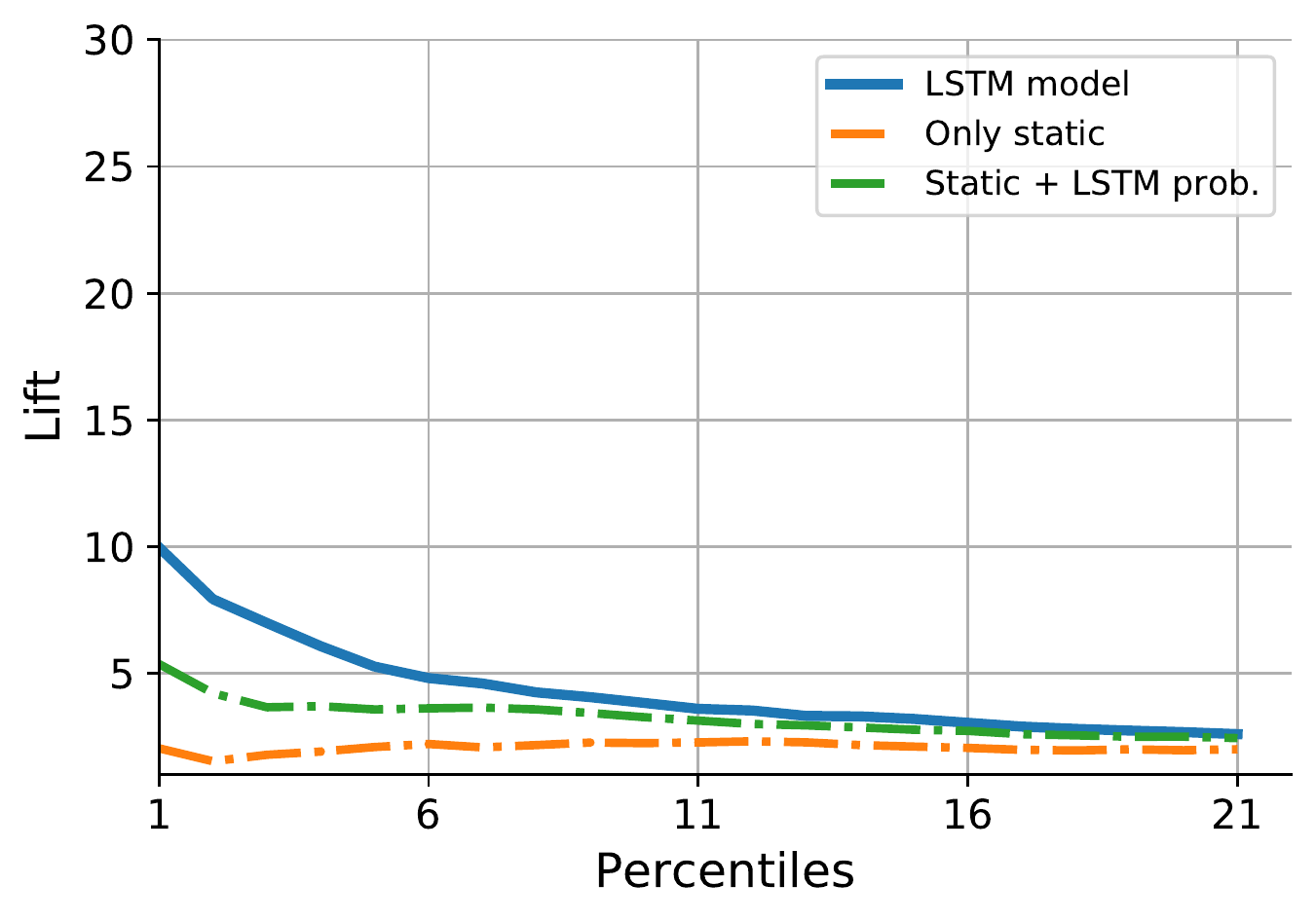}
	\end{subfigure}%
	\begin{subfigure}{0.32\textwidth}
		\subcaption{Outer Fold 3}
		\includegraphics[width=\linewidth]{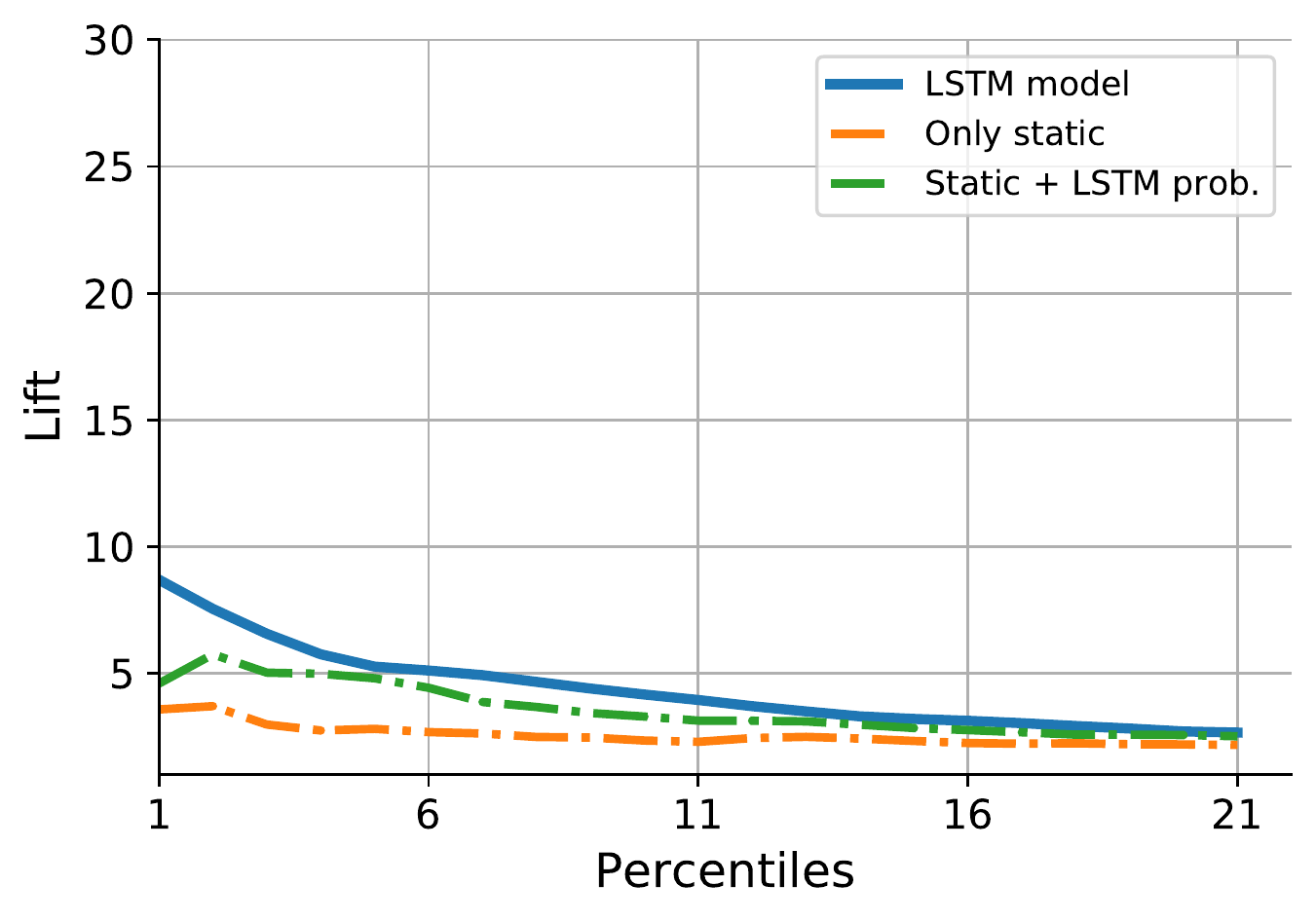}
	\end{subfigure}
	\vspace{-10pt}
	\floatfoot{\footnotesize{\textit{Source:} Own calculations. \\ \textit{Notes:} LSTM refers to model where churn is the target and RFM data the features. Only static refers to l1-regularized logistic regression that uses only customer demographics, customer behavior, and customer contact variables. Static is l1-regularized logistic regression that includes static features as well as the estimated out-of-sample probability from a LSTM model. To estimate the percentiles the data is sorted based on the probabilities of the model from larger to smaller, thus percentile 1 is the 1 percent with the highest estimated churn probability. Figure shows only the first 20 percentiles.}}
\end{figure}

Figure~\ref{fig:lift} shows the lift curves of the first 20 percentiles for each of the outer folds. The figures clearly show that the lift metric of the LSTM model and the logistic model plus fitted probabilities tend to give the best results (a higher lift). 

\newpage
\addcontentsline{toc}{section}{References}
\bibliography{references_cdnn}

\end{document}